\documentclass[sigconf,10pt,authorversion]{acmart}

\usepackage{balance}
\usepackage{booktabs}
\usepackage{framed}
\usepackage{pmboxdraw}
\usepackage{tikz}

\graphicspath{{./figures/}}
\DeclareGraphicsExtensions{.pdf,.jpeg,.png}

\newcommand{\src}[1]{\texttt{\small#1}}

\newcommand{\idea}[1]{
\setlength{\FrameSep}{6pt}
\setlength{\OuterFrameSep}{2pt}
\begin{framed}\noindent #1\end{framed}
}

\copyrightyear{2019}
\acmYear{2019}
\setcopyright{acmlicensed}
\acmConference[Programming '19]{Companion of the 3rd International Conference on Art, Science, and Engineering of Programming}{April 1--4, 2019}{Genova, Italy}
\acmBooktitle{Companion of the 3rd International Conference on Art, Science, and Engineering of Programming (Programming '19), April 1--4, 2019, Genova, Italy}
\acmPrice{15.00}
\acmDOI{10.1145/3328433.3328454}
\acmISBN{978-1-4503-6257-3/19/04}

\begin{document}
\title{Draw This Object: A~Study~of~Debugging~Representations}

\author{Mat\'u\v{s} Sul\'ir}
\orcid{0000-0003-2221-9225}
\affiliation{%
  \institution{Technical University of Ko\v{s}ice}
  \streetaddress{Letn\'a 9}
  \city{Ko\v{s}ice}
  \postcode{042 00}
  \country{Slovakia}
}
\email{matus.sulir@tuke.sk}

\author{J\'an Juh\'ar}
\affiliation{%
  \institution{Technical University of Ko\v{s}ice}
  \streetaddress{Letn\'a 9}
  \city{Ko\v{s}ice}
  \postcode{042 00}
  \country{Slovakia}
}
\email{jan.juhar@tuke.sk}

\begin{abstract}
Domain-specific debugging visualizations try to provide a view of a runtime object tailored to a specific domain and highlighting its important properties. The research in this area has focused mainly on the technical aspects of the creation of such views so far. However, we still lack answers to questions such as what properties of objects are considered important for these visualizations, whether all objects have an appropriate domain-specific view, or what clues could help us to construct these views fully automatically. In this paper, we describe an exploratory study where the participants were asked to inspect runtime states of objects displayed in a traditional debugger and draw ideal domain-specific views of these objects on paper. We describe interesting observations and findings obtained during this study and a preliminary taxonomy of these visualizations.
\end{abstract}

%
%



\maketitle

\section{Introduction}

One of the most important features of any development environment -- whether a live or a traditional one -- is the ability to show the current internal state of the program to the programmer \cite{Victor12learnable}. In contemporary IDEs (integrated development environments) for mainstream programming languages, this feature is incorporated in the debugger, as a part called the \textit{object inspector} or \textit{variables view}.

Traditionally, an expandable tree view of variables in scope is shown. Although this approach is universal, it has its shortcomings. First, the developer can easily suffer from information overload, since non-primitive variables have fields, which in turn have their own sub-fields, etc. Second, this kind of visualization is too generic -- no matter whether the object being displayed is a widget or a network socket, the form is always the same.

For these reasons, approaches and tools for domain-specific object views started to emerge. A domain-specific visualization aims to display each object in a custom way, depicting its essence. As a quick example, an object containing geographical coordinates could be displayed as a point on a map instead of a list of two numbers. Domain-specific visualization approaches range from simple proof-of-concepts such as DoodleDebug \cite{Schwarz11doodledebug} and Vebugger \cite{Rozenberg14templated} to the advanced ecosystem of the Moldable Tools \cite{Chis15moldable}. The mentioned works are primarily concerned with source code frameworks and technical details which should enable the developers to construct their own domain-specific views by scripting. In contrast with them, in this paper we would like to investigate more conceptual questions, such as:

\begin{itemize}
\item In what various ways can states of objects be visually represented and what interesting properties do these views have?
\item Can all objects be meaningfully described by a visualization?
\item What clues could help us to construct domain-specific views without hand-coding them for each class separately?
\end{itemize}

Visualizing a circle is considered easy. But one may wonder: How do I visualize a \src{TCPSocket}? A \src{TextReplacer}? Or a \src{DefaultAdvisorChainFactory}? According to Bret Victor, this is the wrong type of questions to ask and we should design data structures which are easily visualizable instead \cite{Victor12learnable}. Although we agree that programming needs to change and designing visualizable structures is a part of this long process, we strongly oppose the ``wrong question'' part. Trillions of lines of code have been written using traditional programming techniques. Instead of throwing all this effort away, we should strive to gradually improve the state of the art, even if the results might not be ideal.

Therefore, we selected a diverse set of classes from projects written in Java -- a popular but notoriously verbose and rigid language. Participants of our study were asked to draw domain-specific visualizations of the given object instances of our selected classes, displayed in a traditional debugger. In the following sections, we describe the method in detail and summarize our observations and findings.

\section{Method}

Now we will describe the selected projects and classes, the participants of the study and its procedure.

\subsection{Projects}

Our aim was to select projects from diverse domains, relatively easily understandable for newcomers, and easily buildable from source. The following three projects were selected:

\begin{itemize}
\item Java Text Tables\footnote{\url{https://github.com/iNamik/java_text_tables}}: a generator of textual tables using ASCII characters -- e.g., for the use in a console,
\item Nominatim Java API\footnote{\url{https://github.com/jeremiehuchet/nominatim-java-api}}: a client library for a web service which returns geographical coordinates given an address (or vice versa),
\item Chlorine-finder\footnote{\url{https://github.com/dataApps/chlorine-finder}}: a library to detect and redact sensitive data (addresses, e-mails, etc.) in text.
\end{itemize}

\subsection{Classes and Objects}

From the three projects, we chose a set of classes, aiming for the diversity in purpose, complexity, member variables count, and potential for visualization. A total of 11 classes were selected. For their overview, see Table~\ref{t:objects}.

Since the essence of the study was to draw concrete objects at runtime, we had to provide our participants with specific instances of the classes. To arrange this, we selected test cases which covered the given classes. Then we designated breakpoints in the selected classes on places where the objects were already sufficiently initialized. A combination of a test case executed in the debug mode and a breakpoint on a given line thus defined a reproducible runtime state of each object. The resulting list consists of 12 objects, since for the \src{Border} class, we selected two test cases.

\begin{table*}
\caption{A list of classes used in the study} \label{t:objects}
\begin{tabular}{llp{11.3cm}}
\toprule
Project & Class & Description \\ \midrule
Java Text Tables & SimpleTable & A textual table -- an object with a supposedly obvious visual representation. \\ 
~ & GridTable & Similar to SimpleTable, but the final dimensions must be specified during creation. \\
~ & Border & Defines the border style of tables. It contains a member of class Chars with 11 fields. E.g., for a double border, bottomLeft = ``{\small\textSFxxxviii}'', rightIntersect = ``{\small\textSFxxiii}'', etc. \\
~ & LeftAlign & A singleton without instance variables, used in an implementation of function currying (to gradually set parameters, such as a padding character). \\
~ & TopPad & Analogous to LeftAlign. \\ \midrule
Nominatim API & BooleanSerializer & A fieldless class, used as a parameter of Java annotations. It contains a single method named handle, which converts ``true'' or ``false'' into ``1'' or ``0''. \\
~ & Address & A relatively large data class, consisting of 17 instance variables (5 non-primitive), getters and setters. It represents a place on Earth. \\
~ & OsmType & An enumeration (either a NODE, WAY, or RELATION). \\
~ & JsonNominatimClient & A complicated class, something like a central control point of the API. \\ \midrule
Chlorine-finder & Redactor & Combines a finding engine, which searches for sensitive data in a text, with a replacer, which changes them to strings such as ``xx-xx''. \\
~ & MaskFactory & A factory which creates Maskers after loading their configuration from an XML file. A Redactor (above) is one such Masker. \\
\bottomrule
\end{tabular}
\end{table*}

\subsection{Participants}

In total, 33 subjects participated. Thirty of them were master's students of the Evolution of Software Systems course. It is an elective course intended for students interested in programming. According to the post-study survey, 80\% of them have already had industrial programming experience.

The remaining participants consisted of one PhD student and two assistant professors working in the same department as the researchers.

\subsection{Procedure}

First, the subjects read a short introduction to domain-specific visualizations. It included a simple example of a \src{Rectangle} object -- first displayed in a traditional debugger and then as a picture of a rectangle labeled with its dimensions.

Each participant was assigned a project, exceptionally two projects. The subjects were given enough time to become acquainted with the source code, run and inspect unit tests, etc. They were also assigned a selection of 3 objects (instances of 3 classes from Table~\ref{t:objects}).

For each object, they were asked to place a breakpoint on a specified line (e.g., \src{GridTable:73}) and run the given unit test (such as \src{GridTableTest}) in the debug mode. The Variables View window of their IDE then displayed a list of variables in scope, including the ``\src{this}'' variable, which was always of the corresponding class (\src{GridTable} in this case). The task was to draw the ``\src{this}'' object on paper -- in a way which should facilitate program comprehension.

The instructions said the drawings do not have to exhaustively depict all aspects of the objects because the traditional tree-view would remain available in the IDE. Animation and interaction possibilities could be outlined in drawings and further described textually in the web form. Multiple views of the same object could be drawn if necessary.

The master's students were also asked to describe the synopsis of each drawing textually in the web form. For the remaining three participants, a think-aloud protocol \cite{Wohlin12experimentation} was applied: The subjects were speaking out their thoughts and the researchers were taking notes.

One participant decided to skip the drawing of two objects, another two objects could not be displayed in the debugger because of technical problems. Therefore, we received a total of 95 object drawings.

\subsection{Analysis}

All papers were scanned and then inspected by the researchers. During the inspection, both researchers took notes about interesting observations and independently tagged all drawings using qualitative coding \cite{Saldana16coding}, searching for common patterns and distinguishing characteristics in the data. Then they collectively constructed a taxonomy during personal meetings.

The drawings of all participants who agreed to publish them are available online\footnote{\url{https://doi.org/10.6084/m9.figshare.7823774}}.

\section{Observations and Findings}

Now we will describe our observations and findings, grouped by the categories of the taxonomy.

\subsection{Form}

First, we will look at an overall form of the drawings.

\subsubsection{No Visualization}

Classes \src{TopPad} and \src{LeftAlign} from Java Text Tables and \src{BooleanSerializer} from Nominatim API have something in common -- they have no instance variables. Although they may seem like borderline cases, it is fairly common. Often the underlying reason is to utilize polymorphism in various ways, but without a need to store any state.

Interestingly enough, about half of the subjects did not have any idea how to suitably draw objects without instance variables. Their responses ranged from ``I do not know'' and ``N/A'' to drawings of empty rectangles. Some ideas of the other half will be reviewed later in this article.

\subsubsection{Text}

A few participants represented some objects using a text-only notation, such as JSON (JavaScript Object Notation): ``I consider the JSON view more than sufficient,'' said one of them. The use of this notation was observed almost exclusively in data-only objects (\src{Address}).

Since an enumeration (\src{OsmType}) is a simple object, representing it textually was considered suitable by multiple participants. The simplest drawing contains only the enumeration value, as written in the source code: ``NODE''. Other subjects used a combination of the value and the standard string representation (the result of calling the \src{toString} method on this object), in this case, ``NODE: N''.

\subsubsection{Graphics}

A vast majority of the drawings have some kind of a graphical or semi-graphical form. This includes tables, diagrams, and free-form drawings. In the following sections, we will describe our findings regarding these representations.

\subsection{Included Data}

An important question follows: Where is the origin of the data reflected in the drawings?

\subsubsection{Class Names}

The drawings often encompassed the name of the class of the object being drawn. Sometimes they were simplified or otherwise changed: ``table'' instead of \src{SimpleTable}, ``padding top'' instead of \src{TopPad}.

Since \src{BooleanSerializer} completely lacks instance variables, one participant resorted to the visualization of only a static structure of the class. It was a UML-like drawing of the class, including the class name, a list of implemented interfaces, etc.

\subsubsection{Field Names}

\begin{figure}
\includegraphics[scale=1.2]{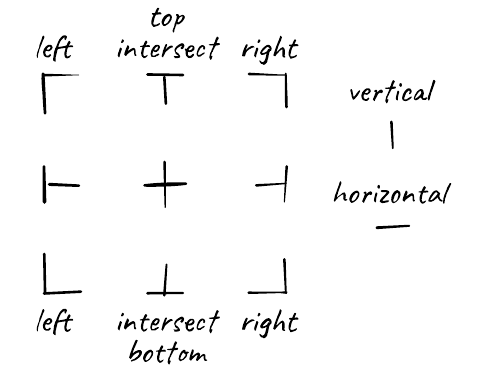}
\caption{A view of an object of class Border with conveniently spaced field values and split field names (The participant's drawing was recreated as a vector image to achieve higher visual quality.)} \label{f:field-names}
\end{figure}

Field names were often displayed as a part of standard ``name: value'' pairs in tables and diagrams. However, we also encountered one particularly creative representation of a \src{Border} object, which we can see in Figure~\ref{f:field-names}. First, all field values were spatially laid out according to their meaning. Then, all two-word field names were split to individual words, which were placed next to the corresponding values. For example, we can see that the value of the \src{topRight} field is ``\scalebox{1.2}[1.7]{\textlnot}''.

Static final fields (constants) were generally considered unimportant when drawing an object state, since their value if not dependent on a specific object instance. However, one participant included the text ``double border'' in a drawing of a \src{Border}. If we wanted to automate this, the only viable data source of the word ``DOUBLE'' would be the name of the constant \src{DOUBLE\_LINE}. In fact, the object the participant was drawing was the value of this constant.

\idea{An object can be checked for identity (or equality) with constants of its type. If a match is found, we can show the given constant name in the visualization.}

\subsubsection{Field Values}

Values of the member variables of the objects being described are almost omnipresent in our responses. In general, field values can be projected into a visualization in various ways, ranging from a literal inclusion of their textual representations to a subtle indication of their characteristics.

If present in the drawings, characters and strings were written verbatim since this is their natural representation. For example, a table cell containing lines ``Left'' and ``Top'' stored as strings was drawn as a rectangle with these strings inside:
\begin{tikzpicture}[baseline=3pt,every node/.style={inner sep=0,outer sep=0}]
\draw (0,0) -| (0.6, 0.4) -| (0,0)
node[pos=0.5,below right=0.5pt]{\tiny\textsf{ Left}}
node[pos=1,above right=0.4pt]{\tiny\textsf{ Top}};
\end{tikzpicture}\,.

Values of numeric fields were sometimes represented literally. For instance, the \src{tableWidth} and \src{tableHeight} members of a \src{GridTable} were written as labels of the rectangle edges in some drawings:
\begin{tikzpicture}[baseline=-3pt,every node/.style={inner sep=0,outer sep=0}]
\draw (0,0) -| (0.3, 0.15)
node[pos=0.25,below=1pt]{\tiny\textsf{4}}
node[pos=0.75,right=1pt]{\tiny\textsf{2}}
-| (0,0);
\end{tikzpicture}. On the other hand, the \src{numRows} and \src{numCols} fields defined in the same class were always expressed only implicitly -- by drawing a grid with the given number of rows and columns:
\begin{tikzpicture}
\draw[xstep=0.2,ystep=0.09] (0,0) grid (0.6,0.27);
\end{tikzpicture}\,.

Booleans were presented in standard ways, such as ``true'' or 1, but also as icons (\checkmark).

One-dimensional arrays and collections were depicted as comma-separated lists or tables.

Arrays of arrays and other types of nested collections were always represented by a grid \begin{tikzpicture}
\draw[step=0.09] (0,0) grid (0.27,0.27);
\end{tikzpicture}
instead of a tree-view which is utilized in many IDEs. However, this might be also affected by the fact that the arrays were not jagged (the number of sub-elements in each element was equal). As a helping factor in the automated selection of an appropriate view, \textit{heuristics} using names could help. For instance, note that our classes representing tables end their name with ``Table''.

Some of the table views encompassed the indexes of all rows and columns. The indexes were even named as ``row'' and ``column'' in one of the drawings (
\begin{tikzpicture}[every node/.style={inner sep=0,outer sep=0}]
\draw (0,0) -| (1.0, 0.15) -| (0,0)
node[pos=0.25,above=1pt]{\tiny\textsf{COLUMN 0}}
node[pos=0.75,left=2pt]{\tiny\textsf{ROW 0}};
\end{tikzpicture}
). If we wanted to automate the dimension naming, we could try to analyze which variables are used in indexing operators, e.g., \src{table[row][col]}.

\idea{Dimensions in multi-dimensional array visualizations can be named after the variables used as indexes.}

\subsubsection{Other Execution Data}

When we think about an ``object state'', we tend to think about its fields and sub-fields. However, we often forget about the current execution stack -- the method being currently executed, its parameters, values of the current local variables, etc. Although they are not traditionally perceived as a part of the object itself, they often contain relevant information about it. For example, although the \src{LeftAlign} class does not have any fields, it has the method \src{apply()} with multiple parameters. In our study, the breakpoint was inside this method, so its parameters were accessible on the stack. Two participants included values of some of these parameters in their drawings: a character used to \src{fill} an empty space in a cell (``\string^'') and a \src{width} of the cell (10).

\idea{Variables on the stack (parameters, local variables) can also be included in the object's visualization, even though they are not a part of the object state per se.}

Some participants included in their drawings data which represented a result of a \textit{future execution}, which is problematic with respect to automation. For example, when the breakpoint was on the first line of the \src{handle} method of a \src{BooleanSerializer} and its parameter was ``false'', one drawing contained only the string ``0'', which should have been its return value. However, it was not yet computed at that point.

\begin{figure}
\includegraphics[scale=1.2]{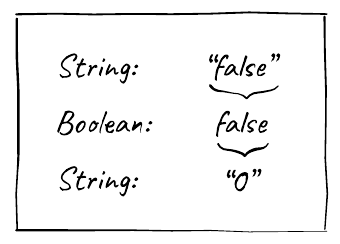}
\caption{A view of a BooleanSerializer, depicting the execution of its method ``handle''} \label{f:process}
\end{figure}

Instead of visualizing the values alone, some subjects tried to express the whole method execution process. In Figure~\ref{f:process}, we can see that the parameter is first converted to a boolean and then to a string.

Some participants were trying to represent the method's behavior in a more general way. In the case of the boolean serializer, instead of capturing only the specific input in the given situation, they expressed multiple valid inputs and their corresponding outputs:
\[
\text{"false"} \rightarrow 0 \qquad \text{"true"} \rightarrow 1
\]

\begin{figure}
\includegraphics[scale=1.2]{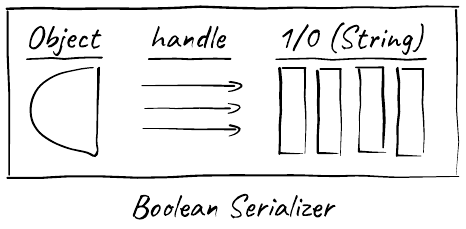}
\caption{A generic BooleanSerializer's behavior view} \label{f:behavior}
\end{figure}

In Figure~\ref{f:behavior}, we can see a graphical and even more generic representation of behavior.

\idea{Objects without fields can be visualized too. One of the possibilities is to depict their behavior.}

\subsubsection{Hypothetical Context}

\begin{figure}
\includegraphics[scale=0.9]{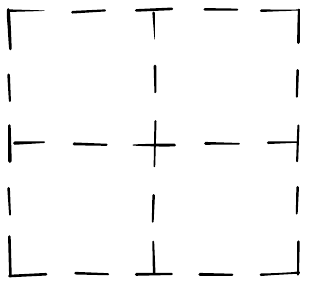}
\caption{A table representing a hypothetical context of the ``Border'' object} \label{f:hypothetical-context}
\end{figure}

When drawing a \src{Border} object, multiple participants drew its context, specifically a table which uses such a border. However, in some cases, it was a 2x2 table -- and a table with such dimensions is never created during the execution of the supplied program. This means that instead of drawing a real context, they designed a hypothetical context -- i.e., a situation which never occurs during the execution of a program, but which is suitable for a presentation of certain features of an object. We can see the discussed visualization in Figure~\ref{f:hypothetical-context}. 

\subsubsection{External Data}
In some cases, the drawings contained information beyond that contained in the source code or execution data. For example, we encountered a map of the world with a point representing the given \src{Address}. Full automation of the generation of similar visualizations is definitely problematic, since it requires domain knowledge and external data source providers.

\subsection{Excluded Data}

Once the number of fields and sub-fields in a given object exceeds some limit, displaying everything would cause information overload. An obvious idea which was suggested by some subjects is to display fields of non-primitive types in a form of a hyperlink, a collapsed tree view, or a similar navigation mechanism. There is nothing bad with occasional use of this approach. However, by applying it uniformly for all fields in all objects, we would achieve nothing more than a traditional Variables View available in current IDEs. Therefore, a better approach is to prefer showing some kinds of data while (temporarily) hiding others, according to some criteria.

\subsubsection{No Exclusions}

Some visualizations drawn by the subjects were complete -- they contained information about all fields (and recursively for their sub-fields) in some sense. We encountered such visualizations only for the enumeration \src{Osm\-Type} and objects of class \src{Border} and \src{GridTable} (counting only classes with fields, of course). When fully expanded, up to the level of primitives and strings, the IDE Variables tree-view of a \src{Border} has 12 items and \src{GridTable} 27 items. However, a large portion of the \src{GridTable}'s fields were empty arrays in our study.

\subsubsection{Empty Arrays}

When drawing a \src{GridTable} which had only the first cell filled with data, about a half of the participants depicted the whole table, while the other half drew only one cell. Therefore, omitting empty arrays or other collections from visualizations may or may not be suitable, depending on a situation.

\subsubsection{Null Values}

In complicated objects such as \src{JsonNomi\-natimClient}, there were many fields with \src{null} values. The subjects tended to hide or collapse such fields in their drawings: ``They shouldn't burden the user for now,'' said one of them.

\idea{Uninitialized variables are suitable candidates for exclusion from visualizations.}

\subsubsection{Implementation Details}

Important adepts for exclusion from visualizations are ``implementation details''. What exactly constitutes an implementation detail is unclear, since private member variables are, by definition, implementation details of an object. One participant excluded internal structure of some fields in a \src{JsonNominatimClient} because they were defined in a third-party library. Although this is definitely not an ultimate rule, it can be a hint for a potential automated visualization generator.

\src{JsonNominatimClient} contains a variable \src{httpClient}, which itself consists of many fields. One participant was looking at the source code to find whether they were supplied to the \src{httpClient} via a constructor, setters or other public methods. According to him, if a client code of an API supplies a value to another class (e.g., configuration settings such as a port number), then it is not an implementation detail, otherwise it could be omitted from the visualization.

\idea{In complicated objects, implementation details can be omitted. They may include variables of types defined in third-party libraries and sub-fields whose values were not supplied as parameters of public API methods.}

\subsubsection{Fields of Given Types}

\begin{figure}
\includegraphics[scale=1.2]{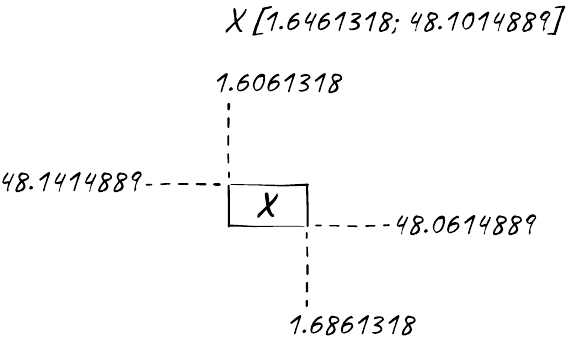}
\caption{A view of an Address showing exclusively its numeric fields} \label{f:numeric}
\end{figure}

Some views focused solely on numeric fields, excluding variables of all other types. In Figure~\ref{f:numeric}, we can see an \src{Address} object displayed as a point with coordinates (X) and as a bounding box, again with exact numeric coordinates listed. In contrast with this representation, another participant depicted address as a table containing the name of the settlement, county, state, and other purely string-typed fields.

\subsection{Relation Handling}

In the following sections, we will discuss what kinds of relations between objects the participants perceived and how they depicted them.

\subsubsection{Grouping}

The \src{JsonNominatimClient} contains multiple similarly named fields, such as \src{searchUrl}, \src{reverseUrl}, and \src{lookupUrl}. In a drawing, a participant visually grouped them into a separate table named ``URLs'' with rows named ``search'', ``reverse'' and ``lookup''.

\begin{figure*}
\includegraphics[scale=1.2]{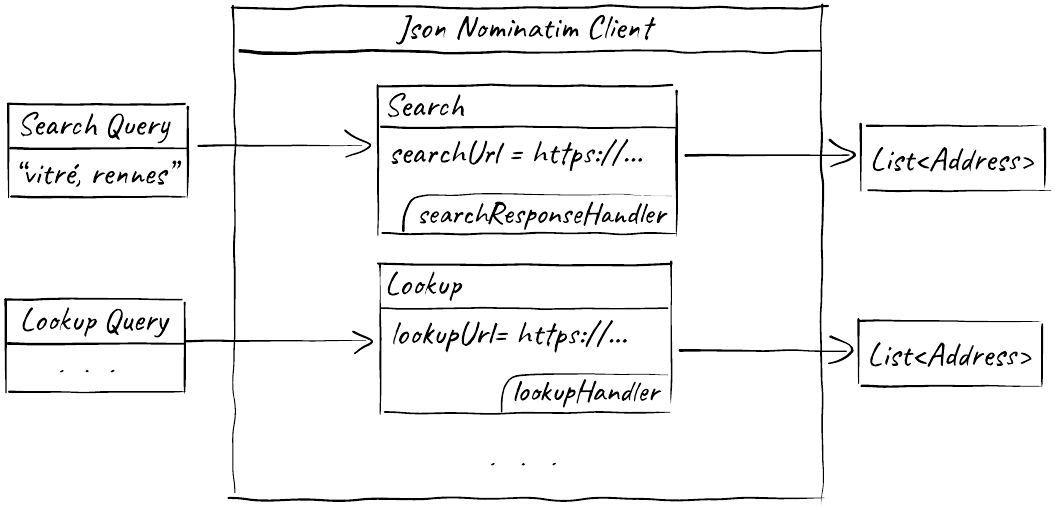}
\caption{A JsonNominatimClient view grouping search- and lookup-related variables (simplified by researchers)} \label{f:grouping}
\end{figure*}

Interestingly enough, another participant also grouped the fields, but in a different way. He drew a box for each request type (``search'', ``lookup'', ``reverse'') and included variables related to each of them in the corresponding box. For example, in the first box, there was a \src{searchUrl} and a ``search response handler''. We can see this view in Figure~\ref{f:grouping}.

We also encountered the grouping of variables by their runtime values. Specifically, all values with null variables were grouped at the bottom of the table, in an expandable row named ``other attributes'' with a value ``all null''.

\idea{Fields can be grouped by their name, purpose or value.}

\subsubsection{Nesting}

\begin{figure}
\includegraphics[scale=1.2]{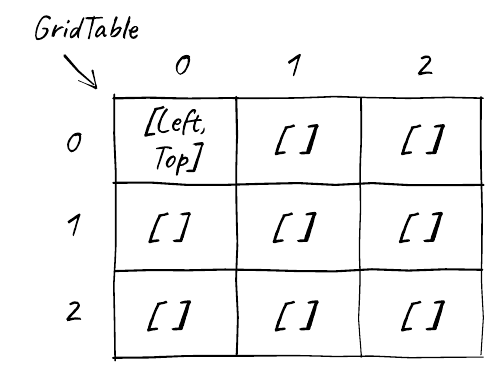}
\caption{A textual array view nested in a graphical GridTable} \label{f:nesting}
\end{figure}

In Figure~\ref{f:nesting}, we can see a visual representation of a 3x3 \src{GridTable}. Therefore, inside the cells, we would expect their visual content: the text ``Left Top'' in the first cell and an empty space in the rest of them. Instead, we see debugging string representations of Java collections, such as \src{[Left, Top]} or \src{[]}. Therefore, various representations can be nested inside each other. In our study, we have even encountered a drawing with multiple levels of nesting.

\subsubsection{Connecting}

\begin{figure*}
\includegraphics[scale=1.2]{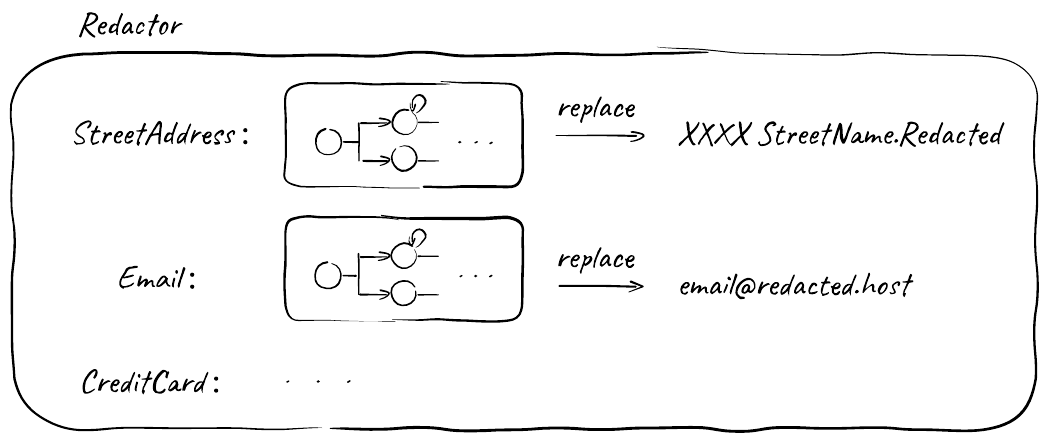}
\caption{Patterns and replacements of a Redactor, joined by their names} \label{f:connecting}
\end{figure*}

Relationships between two object tended to be drawn using common connectors, such as lines or arrows. Except for traditional memory references between objects, we also noted \textit{conceptual references}. For example, a \src{Redactor} contains a list of \src{finders} (\src{name}: ``Email, \src{pattern}: ``\textbackslash{}b[A-Z]\dots'') and a hash map named \src{replacements} (``Email'' $\rightarrow$ ``email@redacted.host'', \dots). A drawing in Figure~\ref{f:connecting} shows that the finders and replacements were ``joined'' by their names (e.g., ``Email'') into triples of the form name--pattern--replacement. In the same figure, we can also see an example of a railroad diagram representing the regular expression nested in the overall view of this object.

Similarly, an \src{Address} object contained a \src{String}-typed field named \src{osmType} with the value ``node''. Although this field was not of the type \src{OsmType}, a participant drew an arrow between it and the enumeration \src{OsmType.NODE}.

\idea{Conceptual references between objects can be included in visualizations -- e.g., two objects can be ``joined'' by a common value of one of their fields.}

\subsection{Size}

Visualizations of different sizes and complexities have distinctive potential use cases in IDEs.

\subsubsection{Space-Constrained Representations}

Some participants drew simple, \textit{icon}-like drawings. For instance, the most succinct representation of a \src{Border} in our study was a mere horizontal double line drawing: ``\scalebox{2}[1]{=}'', i.e., a value of its field \src{horizontal}. Although it does not express all necessary details, such terseness can be useful in situations when space is constrained. This includes visual source code augmentation \cite{Sulir18visual}, such as our recent approach for displaying sample variable values directly at the end of each line in a source code editor \cite{Sulir18augmenting}. It can also help a developer to improve orientation and navigation when a lot of objects are presented at the same time -- which is a purpose of the icon indeed. In our study, we have also spotted an elegant icon representing a \src{TopPad} object
\begin{tikzpicture}[baseline=1pt]
\draw[fill=lightgray] (0,0) rectangle (0.35,0.35);
\draw[fill=white] (0.09,0.09) rectangle (0.26,0.25);
\draw[fill=darkgray] (0,0.25) rectangle (0.35,0.35);
\end{tikzpicture}
and an \src{OsmType} with the value \src{NODE}:
\begin{tikzpicture}[baseline=1pt]
\draw (0,0) rectangle (0.35,0.35)
node[midway,circle,fill,scale=0.4] {};
\end{tikzpicture}\,.
\idea{A simpler object can be often succinctly visualized by an icon representing its class, enumeration value, or one of its field values.}

In a tree-like diagram of the \src{Address} object, there were icons for numeric types
\begin{tikzpicture}[baseline=2pt]
\draw (0,0) rectangle (0.45,0.34)
node[midway] {\scriptsize\textsf{\textbf{123}}};
\end{tikzpicture}
and strings
\begin{tikzpicture}[baseline=2pt]
\draw (0,0) rectangle (0.50,0.34)
node[midway] {\scriptsize\textsf{\textbf{Abc}}};
\end{tikzpicture}.
In these cases, icons were only supplementing the views of actual values instead of trying to replace them.

\subsubsection{Standard-Sized Views}

A majority of visualizations designed by our participants were too large to be used as a word-size graphic or icon. However, they would fit into auxiliary IDE tool windows displayed alongside the source code and other views. Drawings from Figures~\ref{f:field-names}--\ref{f:numeric} and \ref{f:nesting} belong to this category.

\subsubsection{Large Diagrams}

Particularly for complicated objects like a \src{JsonNominatimClient} or a \src{Redactor}, we encountered large diagrammatic drawings which barely fit the A4 paper used during the study. The most sensible application of such visualizations would be to display them in a full-screen mode.

\subsection{Interaction} \label{s:interaction}

Although pen and paper are not a perfect medium to express interaction, multiple participants described interaction possibilities in their written or spoken notes. A vast majority of them aimed to suppress potential information overload.

\subsubsection{Expansion}

When the number of fields to display was too large, the subjects included controls to expand and collapse details in their visualizations. Expansion controls were present in \src{Address}, \src{JsonNominatimClient}, and \src{Redactor}. A common representation was a familiar plus sign
\begin{tikzpicture}[every node/.style={inner sep=0,outer sep=0}]
\draw[gray,thick] (0,0) rectangle (0.25,0.25)
node[midway,gray] {$\small\boldsymbol{+}$};
\end{tikzpicture}\,. After expansion, a nested view of a given field should be displayed.

\subsubsection{Hyperlinks}

Some objects consist of a collection of otherwise relatively independent objects. For example, a \src{MaskFacto\-ry} can contain multiple \src{Masker}s. A participant drew it as a table of names and hyperlinks to individual objects:
\begin{tikzpicture}[baseline=1pt,every node/.style={inner sep=0,outer sep=0}]
\draw (0,0) rectangle (1.3, 0.32)
node[midway]{\footnotesize\textsf{``redactor''}};
\draw (1.3,0) rectangle (3.1, 0.32)
node[midway,below=1.4pt]{\scriptsize\textsf{\underline{\smash{Redactor@1656}}}};
\end{tikzpicture}\,.
Our subjects suggested opening each such masker in a separate pop-up window. However, we suggest using alternative interfaces such as a horizontal boundless tape \cite{Taeumel12vivide} or Object Pager \cite{Chis15moldable} to prevent the workspace from being flooded by too many windows.

\subsubsection{Zooming}

One of the subjects suggested a zooming interface \cite{Cockburn09review} may be appropriate to inspect either one cell of a \src{GridTable} or the whole table. We add that zooming interfaces could be also used to inspect runtime objects with various levels of detail -- for example, at a lower zoom, to show only the most important properties of the whole object, and to show details of a specific aspect at a higher zoom.

\section{Threats to Validity}

We used only 11 classes in this study, which can hardly be a representative sample of all the code ever written. Furthermore, the classes were chosen subjectively by researchers. Ideal visualizations highly depend on a particular domain -- and since we studied only 3 projects, the domain selection was limited and relatively random. Nevertheless, we tried to select a diverse set of classes from various points of views to ensure the richness of data, which is an important factor for qualitative studies.

Although we included some deeply nested objects (e.g., \src{JsonNominatimClient}), the majority of objects was rather shallow. Furthermore, inherently graph-based objects were not inspected.

The number of collected responses per class varied from 3 to 16. However, since we derived no important quantitative measures from the responses, we do not consider this a large threat to validity.

Although the participants were not already familiar with the given projects, they were small enough to sufficiently understand in one session (max. 1,600 lines of Java code).

During our study, the participants were not given any specific task to solve. We could obtain more interesting insights by asking them to fix a bug or implement a new feature. Since it may be difficult to say what is a useful visualization without a task specified, studies of task-oriented visualizations are an important future research direction.

The observations and findings described in this paper might not be generalizable, which is a common shortcoming of many qualitative studies. However, we perceive this article mainly as a collection of ideas and a starting point for further research.

\begin{figure*}
\includegraphics[scale=0.9]{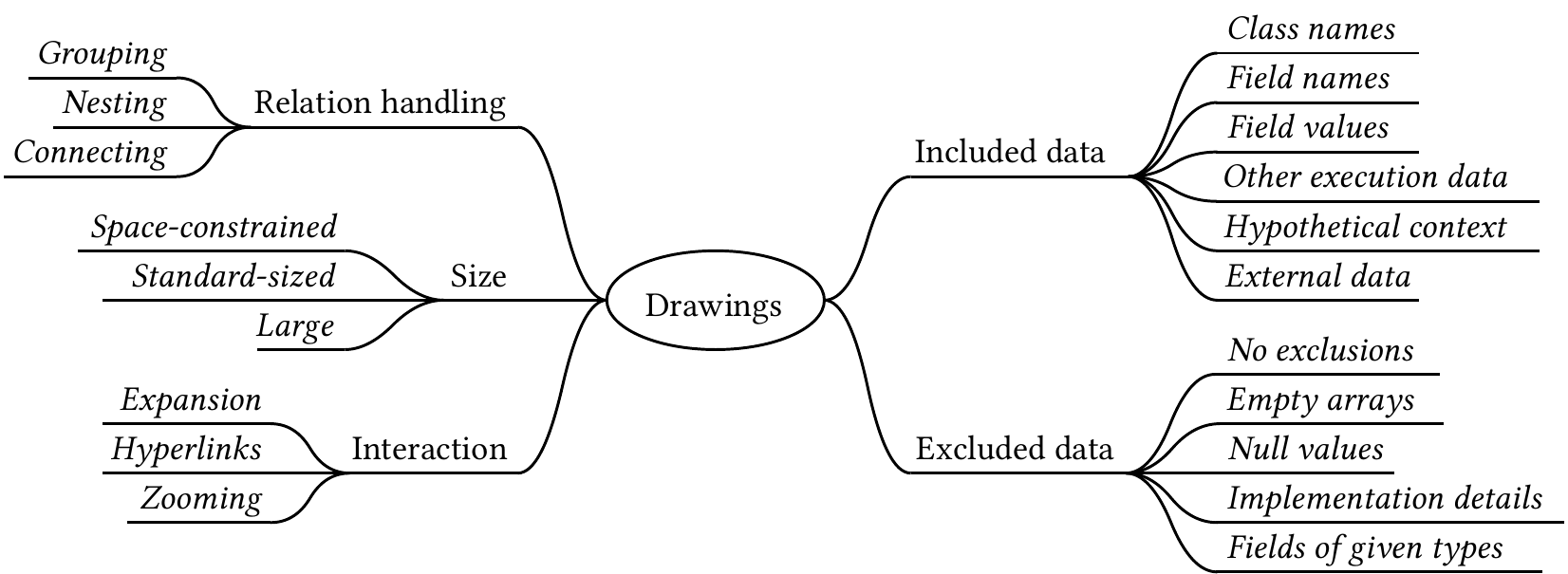}
\caption{The taxonomy of drawings of runtime objects} \label{f:taxonomy}
\end{figure*}

\section{Related Work}

Software visualization is a vast research area, ranging from static source code visualization \cite{Cruz09code} to algorithm animation \cite{Simonak13algorithm}. Therefore, we will focus on viewers of runtime states of objects, particularly the ones displaying one state at a time. Then we will review related empirical studies involving visual object representations or drawings in general.

\subsection{Visualization Approaches}

On one end of the spectrum, there are visualizations which are usable in any context and for all types of objects, but which are very generic and may not be optimal for high-level comprehension. DDD \cite{Zeller96ddd}, JIVE \cite{Gestwicki05methodology}, and HDPV \cite{Sundararaman08hdpv} pertain to this category.

Some tools are focused only on specific contexts or data types, which enables them to display high-level views fully automatically, i.e., without any additional scripting. For example, a tool by Alsallakh et al. \cite{Alsallakh12visualizing} shows searchable lists, line charts and frequency-based visualizations for arrays. Hoffswell et al. \cite{Hoffswell18augmenting} designed a set of sparklines (word-sized visualizations) for selected data types, such as numbers and sets. Kanon \cite{Oka17live} displays node-link diagrams for data structures, particularly linked lists, in a live programming environment. Reactive Inspector \cite{Salvaneschi16debugging} is limited to reactive programs.

In scriptable approaches, source code must be written to produce a given visualization. Each script is applicable in a given context -- usually for a given class. DoodleDebug \cite{Schwarz11doodledebug} is based on an idea that we should define graphical representations of Smalltalk objects (in methods \src{drawOn:} and \src{drawSmallOn:}) in the same way as we define the string representations (\src{printOn:}). Vebugger \cite{Rozenberg14templated} suggests a similar practice for Java.

The Moldable Inspector \cite{Chis15moldable} enables construction of domain-specific debugging object visualizations. They are written using the supplied framework which aims to make the creation of new views as simple as possible. The views can include interaction possibilities.

Rein et al. \cite{Rein17exploratory} designed an approach using the Vivide environment to construct custom live views, such as tables or histograms, during the development of data-intensive programs.

Although not aimed directly at debugging, the Morphic user interface framework \cite{Maloney95morphic} supports visualization of selected object properties. The visualization is live, i.e., the visual representation changes when the underlying object is modified.

\subsection{Empirical Studies}

Chi\c{s} et al. \cite{Chis15moldable} asked developers about their general attitude toward object inspectors. They found that the developers would like to display different views of the same object based on the task begin performed and explore logical connections between objects. They also listed 8 categories based on the analysis of the views written for the Moldable Inspector framework. However, the analysis was focused purely on the implementation aspect (they distinguished categories such as a list, tree, Morph, Roassal view). Furthermore,  they did not discuss challenges, such as how to display objects of abstract nature. According to Chi\c{s} et al. \cite{Chis15moldable}, fully automated generation of domain-specific views is an open question, which is one of the topics discussed in our article.

Lee et al. \cite{Lee08how} interviewed programmers about the importance of features in diagramming tools. Among other findings, object diagrams representing runtime states were considered more important than static class diagrams. No further analysis was performed though.

There exist multiple studies focused on developers' drawings. For example, Dekel and Herbsleb \cite{Dekel07notation} studied the notation used in design drawings. Baltes and Diehl \cite{Baltes14sketches} investigated the use of sketches and diagrams in practice. None of these studies is focused on runtime views or object state representations.

\section{Conclusion and Future Work}

We presented an exploratory study in which the participants were asked to draw the given objects in the program on paper. The perception of a suitable visual representation varied greatly across both objects and individuals. After describing the taxonomy of the drawings in detail, we now present it visually in Figure~\ref{f:taxonomy}.

For every object in our study, we received at least a few meaningful drawings. Although we cannot draw final conclusions, we are inclined toward the idea that all objects can be visualized in a domain-specific way. However, the more complex or abstract is an object, the more difficult and human-intensive process is necessary to produce its visualizations. Therefore, full automation may be unattainable for all objects. Furthermore, for some objects, the view of the process in which the object is involved is more important than its state. In spite of that, we described multiple ideas which can either partially improve the current fully automated generic views, or ease the manual creation of domain-specific views.

To summarize, our contributions are:
\begin{itemize}
\item a method to brainstorm possible object visualizations without actually implementing them,
\item a dataset of drawings obtained during our study, which can be used for further analysis,
\item a set of ideas to improve both generic (automated) and domain-specific views,
\item a preliminary taxonomy of runtime views of objects.
\end{itemize}

Although in this paper we only touched the surface of the problems regarding the automation of domain-specific view generation, it is a very promising future research direction. On a sample of object--view pairs, we can try to map individual object fields and other available execution data onto the individual parts of the graphical views. Then we could devise heuristics regarding this mapping -- for example, conditions when a numeric field is printed literally and when it is reflected in geometrical properties, such as a length of a line.

In the more distant future, we envision debugging tools which will utilize these heuristics and automatically display at least partially domain-specific object visualizations, even in the absence of any hand-coded representations. Of course, this will be only applicable to a certain degree and in many cases, generic visualizations will be displayed instead.

Our study can be also expanded to include drawings of changes between two or multiple states instead of describing only one state at a time. Since the task being solved can affect the usefulness of visualizations, we should perform studies with specific tasks given to participants. Other contextual information, such as the programmer's knowledge profile \cite{Pietrikova16towards}, could be taken into account. Finding whether domain-specific visualizations designed by one person are comprehensible by other developers is another promising future research idea. Finally, we could try to explore the applicability of the results in areas broader than debugging, such as software reengineering, teaching, or even psychology.

\begin{acks}
This work was supported by FEI TUKE Grant no. FEI-2018-57 ``Representation of object states in a program facilitating its comprehension''. This work was also supported by Project VEGA No. 1/0762/19 Interactive pattern-driven language development.
\end{acks}

\bibliographystyle{ACM-Reference-Format}
\bibliography{px}
\balance

\end{document}